\documentclass[a4]{elsart5p}

\usepackage{amssymb,amsmath} 
\usepackage{bm,hyphenat,xspace} 
\usepackage{graphicx,epsfig}
\usepackage{tabularx}
\usepackage{epsfig} 

\newcommand{\email}[1]{\ead{#1}}

\newcommand {\mbf}[1]{{\mathbf{#1}}}
\newcommand {\vecg}[1]{\mbox{\boldmath{$#1$}} }

\newcommand{\fm}{\;\mathrm{fm}}
\newcommand{\cm}{\mathrm{c\!\:\!.m\!\:\!.}}
\newcommand{\A}[2]{{}^{#1}\mathrm{#2}}
\newcommand{\Ax}[1]{{}^{#1}\mathrm{Z}}

\begin{document}

\begin{frontmatter}

\title {Nonlocal optical potential with core excitation in 
${}^{10}\mathrm{Be}(d,p){}^{11}\mathrm{Be}$ and 
${}^{11}\mathrm{Be}(p,d){}^{10}\mathrm{Be}$ reactions
}

\author{A.~Deltuva}
\email{arnoldas.deltuva@tfai.vu.lt}, 
\author{D. Jur\v{c}iukonis}
\email{darius.jurciukonis@tfai.vu.lt}

\address{
Institute of Theoretical Physics and Astronomy, 
Vilnius University, Saul\.etekio al. 3, LT-10257 Vilnius, Lithuania}



\begin{abstract}
  We propose a new nonlocal form of the nucleon-nucleus optical potential and demonstrate its reliability.
  We extend the nonlocal potential to include the excitation of  the nuclear core
  and develop  energy-independent proton-${}^{10}\mathrm{Be}$ potential reasonably reproducing the
  experimental data at low energies. We  apply the new potential 
  to the study of deuteron stripping and pickup reactions
${}^{10}\mathrm{Be}(d,p){}^{11}\mathrm{Be}$ and ${}^{11}\mathrm{Be}(p,d){}^{10}\mathrm{Be}$
  using rigorous three-body   Faddeev-type equations for transition operators
  that are solved in the momentum-space partial-wave framework.
  The achieved description of the experimental data  is considerably more successful as
  compared to previous studies with local potentials.
  The values of spectroscopic factors consistent with the data are determined, exhibiting only
  weak energy dependence. 
  The results possibly indicate an increased predicting power of the proposed calculational scheme.
\end{abstract}


\end{frontmatter}


\section{Introduction \label{sec:intro}}

Deuteron pickup and stripping  reactions $\Ax{A+1}(p,d)\Ax{A}$ and $\Ax{A}(d,p)\Ax{A+1}$ 
 are extensively used to extract the information on structure of the involved nucleus $\Ax{A+1}$,
that can be either stable such as $\A{17}{O}$ or an exotic short-living nucleus such as $\A{11}{Be}$.
Since these reactions are dominated by three-body degrees of freedom, their theoretical
calculations rely on three-body approaches for the system consisting of a proton ($p$), a neutron ($n$),
and a nuclear core $\Ax{A}$.
The simplest ones and therefore most widely used are the distorted-wave Born approximation (DWBA) 
and adiabatic distorted-wave approximation (ADWA) reviewed in Ref.~\cite{timofeyuk:20a}.
Typical dynamics, mainly due to the calculational
simplicity, is assumed to be determined by local potentials between nucleons and the nuclear core, 
either real (for bound states) or complex (for scattering states).
The uncertainties due to an approximate solution of the three-body problem have been overcome
by rigorous three-body Faddeev scattering theory \cite{faddeev:60a} in the version of integral equations
for the transition operators proposed by Alt, Grassberger and Sandhas \cite{alt:67a}.
Since it was practically implemented in the momentum-space framework \cite{deltuva:09b},
it was possible to include the potentials that are nonlocal in the coordinate space.
Those pioneering calculations \cite{deltuva:09b} revealed important effects of the
optical potential nonlocality in $\Ax{A+1}(p,d)\Ax{A}$ and $\Ax{A}(d,p)\Ax{A+1}$ reactions 
and thereby triggered the developments of the DWBA and ADWA approaches to include nonlocal
optical potentials \cite{timofeyuk:13b,titus:16a}.
The ADWA extension \cite{timofeyuk:13b,PhysRevLett.117.162502} claimed even more spectacular
nonlocality effects like an enhancement of the differential cross section by a factor of two
in particular cases and induced strong sensitivity to the neutron-proton potential model \cite{PhysRevLett.117.162502}.
However,  more refined Faddeev-type \cite{deltuva:18c}
and continuum-discretized coupled-channel (CDCC) calculations \cite{gomez:nlop}
have not confirmed those findings, indicating unreliability of the ADWA in the context
of nonlocal optical potentials. 

Another important dynamic ingredient in few-body nuclear reactions is the excitation of the
nuclear core $\Ax{A}$. In standard reaction approaches the core excitation (CeX) as well as other
inelastic processes are accounted for implicitly via optical potentials. Explicit inclusion of the CeX
generates coupling potentials and states with several components.
When the deuteron stripping and pickup  reactions are analyzed using standard
DWBA and ADWA approaches, the different components are assumed to participate
in the reaction independently, and  the differential cross section for a given transfer channel simply scales with
 the respective spectroscopic factor (SF),
the weight of the respective component in the bound  core plus neutron system.
However, the Faddeev-type calculations extended for the dynamical
excitation of the nuclear core \cite{deltuva:13d} 
proved for a number of cases like ${}^{10}\mathrm{Be}(d,p){}^{11}\mathrm{Be}$ and
${}^{20}\mathrm{O}(d,p){}^{21}\mathrm{O}$
that the SF factorization in the differential cross section 
is a good approximation at low energies only, roughly up to  10 MeV per nucleon,
while the deviation from the SF factorization assumption increases with increasing reaction energy.
Furthermore, the direction of the quadrupole CeX effect depends on the angular momentum transfer $\ell$,
suppressing $\ell=0$ and enhancing $\ell=2$ reactions \cite{deltuva:17b}.
The CeX effect is a result of multi-step transitions leading to 
a complicated interplay between its  contributions of the two- and three-body nature.
Consistent conclusions have been drawn also in the CDCC-type study \cite{gomez:17a}.

Those Faddeev-type calculations with CeX were limited so far to local potentials between nucleons
and the core. A further shortcoming in the case of 
${}^{10}\mathrm{Be}(d,p){}^{11}\mathrm{Be}$ and ${}^{11}\mathrm{Be}(p,d){}^{10}\mathrm{Be}$ reactions
was the use of standard optical potential parametrizations such as Chapel Hill 89 (CH89) \cite{CH89}
or Koning and Delaroche (KD) \cite{koning}, fitted to heavier nuclei.
Those parametrizations do not provide  accurate description of $p+{}^{10}\mathrm{Be}$ scattering
data, examples for elastic scattering at 6 to 11 MeV are shown in Ref.~\cite{dBe12-21c}.
Furthermore, as the parameters of local optical potentials are energy-dependent, there is an ambiguity
in fixing them, especially in transfer reactions where the effective nucleon-core energy differs in the
initial and final states.
Thus, the predictions of previous calculations for three-body reactions have the associated uncertainties.
Therefore, the goals of the present work are (i) developing an energy-independent nonlocal optical potential
for the nucleon-${}^{10}\mathrm{Be}$ system including explicitly the dynamical excitation
of the ${}^{10}\mathrm{Be}$  core, consistent with the experimental data in a possibly broader
energy regime, and (ii) applying this new potential to the description of 
${}^{10}\mathrm{Be}(d,p){}^{11}\mathrm{Be}$ and ${}^{11}\mathrm{Be}(p,d){}^{10}\mathrm{Be}$ reactions
using the rigorous three-body Faddeev formalism. Given their sophistication and advantages over
widely employed standard potentials and approaches, 
the combination of the two above goals would constitute a new
state-of-the-art description of deuteron stripping and pickup  reactions.

Section \ref{sec:eq} introduces the Faddeev equations with the excitation of the core while
Section \ref{sec:pot} describes and validates the proposed nonlocal optical potential.
Section \ref{sec:res} presents the results for ${}^{10}\mathrm{Be}(d,p){}^{11}\mathrm{Be}$
and ${}^{11}\mathrm{Be}(p,d){}^{10}\mathrm{Be}$ reactions at several energy values, whereas the conclusions are
drawn in Sec.~\ref{sec:sum}.
Note that we use natural units $\hbar=c=1$.

\section{Three-body AGS equations with core excitation \label{sec:eq}}

We consider the $p+n+\Ax{A}$ three-particle system with masses $m_p$, $m_n$, and $m_A$, respectively,
in its center-off-mass (c.m.) frame.
The nuclear core $\Ax{A}$ can be either in its ground (g) or excited (x) state, with the
respective components of the operators being indicated 
by Latin superscripts in the following. The transitions between the internal states of the core are
induced by nucleon-core  potentials $V_{\alpha}^{ba}$, where the Greek subscript $\alpha$
labels the spectator particle in the odd-man-out notation, e.g., the spectator $n$ means
the interacting  $\Ax{A}+p$ pair, and so on.
For each interacting pair the two-particle transition operator is obtained
from the Lippmann-Schwinger equation 
\begin{equation}  \label{eq:Tg}
T_{\alpha}^{ba} =  V_{\alpha}^{ba} +\sum_{j=g,x} 
V_{\alpha}^{bj} G_0^{j} T_{\alpha}^{ja}
\end{equation}
where the free resolvent
$G_0^{j} = (E+i0-\delta_{jx}\Delta m_A - K)^{-1}$
has contributions not only from the  kinetic energy operator $K$ but also
from the core excitation energy $\Delta m_A$ in the respective sector of the  Hilbert space,
$E$ being the available energy.

Dynamic equations to be solved  in the present work are the
Faddeev equations \cite{faddeev:60a} in the AGS version  \cite{alt:67a}
for three-body transition operators, extended  in Ref.~\cite{deltuva:13d} to include
 CeX via coupling of different sectors in the Hilbert space, i.e.,
\begin{equation}  \label{eq:Uba}
U_{\beta \alpha}^{ba}  = \bar{\delta}_{\beta\alpha} \, \delta_{ba} {G^{a}_{0}}^{-1}  +
\sum_{\sigma=p,n,A} \, \sum_{j=g,x}   \bar{\delta}_{\beta \sigma} \, T_{\sigma}^{bj}  \,
G_{0}^{j} U_{\sigma \alpha}^{ja},
\end{equation}
with  $\bar{\delta}_{\beta\alpha} = 1 - \delta_{\beta\alpha}$.

On-shell matrix elements
of $U_{\beta \alpha}^{ba}$ taken between the two-cluster  channel states
determine the physical transition amplitudes for the respective reactions.
E.g., for the deuteron induced reactions the initial channel state
$|{\nu_A}, \mbf{q}_A \rangle$  is a product of the  deuteron wave function and a free wave for
the deuteron-nucleus motion with the relative momentum $q_A$, the set of discrete quantum numbers being abbreviated by $\nu_A$.
In the  $p+\Ax{A}$ channel with  the relative proton-nucleus momentum $\mbf{q}_p$
the channel state $|{\nu_p}, \mbf{q}_p \rangle = 
|\nu_p^g, \mbf{q}_p \rangle + |\nu_p^x, \mbf{q}_p \rangle$
has two components whose norms are given by the respective spectroscopic factors,
their summ being equal to unity.
Thus, the amplitudes for the deuteron stripping reaction are
\begin{equation} \label{eq:ampl}
\mathcal{T}(\nu_p,\mbf{q}_p;\nu_A,\mbf{q}_A) =
\sum_{b=g,x} \langle \nu_p^b, \mbf{q}_p| U_{p A}^{bg}  |\nu_A^g, \mbf{q}_A \rangle.
\end{equation}
and the corresponding differential cross sections are
\begin{gather}  \label{eq:d3s}
\frac{d\sigma(\nu_A\to\nu_p)}{d\Omega_p}
= (2\pi)^4 \, M_A M_p \, \frac{q_p}{q_A} \, 
|\mathcal{T}(\nu_p,\mbf{q}_p;\nu_A,\mbf{q}_A)|^2,
\end{gather}
where $M_\alpha$ is the spectator-pair reduced mass for the  partition $\alpha$.
For unpolarized observables one has to perform a corresponding spin averaging (summation) over the initial
(final) states.

The solution of the scattering  equations (\ref{eq:Uba}) is performed  in the momentum-space 
partial-wave framework, with technical details explained in  Refs.~\cite{deltuva:13d,deltuva:17b}.

 \section{Nonlocal potential with core excitation \label{sec:pot}}

 Since our Faddeev-type calculations are performed in the momentum-space basis, the treatment of nonlocal potentials,
 once they are transformed to momentum space, is the same as of local ones. The most often used
 nonlocal form in the coordinate space is the one proposed by Perey and Buck \cite{pereybuck}, i.e.,
 \begin{equation}  \label{eq:Vpb}
V_N(\mbf{r}',\mbf{r}) =  H(|\mbf{r}'-\mbf{r}|) V(y),
\end{equation}
 where 
 $y = |\mbf{r}'+\mbf{r}|/2$, and 
  \begin{equation}  \label{eq:Hx}
 H({x}) = \pi^{-3/2} \rho^{-3} e^{-(x/\rho)^2}
  \end{equation}
 is the nonlocality function with the nonlocality range $\rho$.
 The local part
 \begin{equation}  \label{eq:Vy}
   \begin{split}
   V(y) = {}& -V_V \,f_V(y) - iW_V \,f_W(y) - i 4 W_S\,f_S(y)[1-f_S(y)] \\ {}&   
   + V_s \frac{2}{y} \frac{df_s(y)}{dy} \, \vecg{\sigma}\cdot \mbf{L}
   \end{split}
 \end{equation}
 has real volume, imaginary volume and surface, and real spin-orbit terms with the respective
 strength parameters $V_V$, $W_V$, $W_S$, and  $V_s$, while the
 radial dependence is given by the corresponding Woods-Saxon functions
 \begin{equation}  \label{eq:fws}
   f_k(y) =  [1+ e^{(y-R_k)/a_k}]^{-1}
\end{equation}
with radius $R_k = r_k A^{1/3}$ and diffuseness $a_k$.
 This type of potential, with parameters of Giannini et al. \cite{giannini,giannini2},
 has been used in a number of studies \cite{deltuva:09b,deltuva:18c,deltuva:16d}, demonstrating
 important nonlocality effects in nucleon transfer reactions.

 Perey and Buck \cite{pereybuck} proposed also an alternative version of the nonlocal potential,
 given by Eq.~(\ref{eq:Vpb}) but with  $y = (r'+r)/2$. We make one step further and introduce the
 nonlocal potential
 \begin{equation}  \label{eq:Vdj}
V_N(\mbf{r}',\mbf{r}) =  \frac12 \big[ H(|\mbf{r}'-\mbf{r}|) V(r) + V(r')H(|\mbf{r}'-\mbf{r}|) \big].
\end{equation}
 By expanding in Taylor series it is an easy exercise to show that the leading term in the difference between the
 two latter versions is  $\frac14 H(|\mbf{r}'-\mbf{r}|) V^{(2)}(r) (r'-r)^2$, i.e., a small quantity.

\begin{figure}[!]
\begin{center}
\includegraphics[scale=0.69]{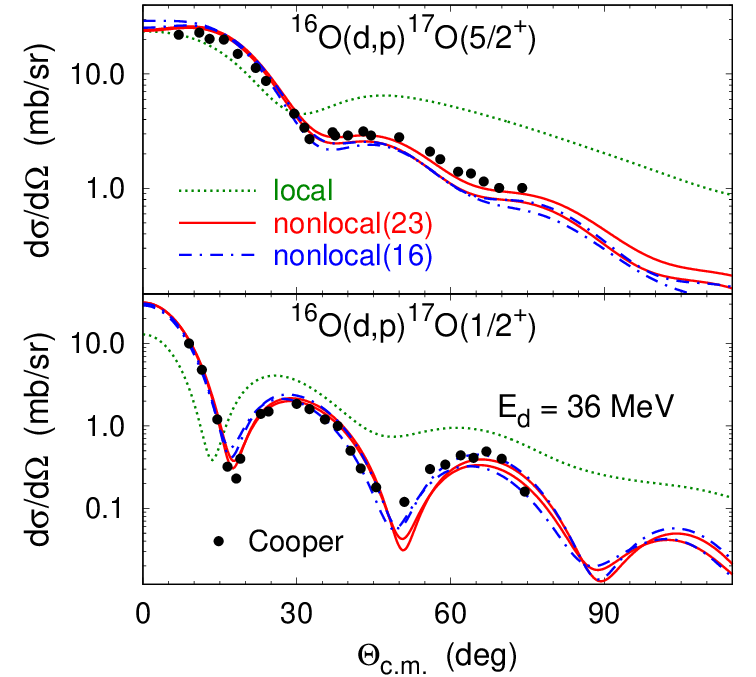}
\end{center}
\caption{\label{fig:O}  (Color online)
Differential cross section  for  $\A{16}{O}(d,p)\A{17}{O}$ transfer reactions
at  $E_d = 36$ MeV leading to $\A{17}{O}$ ground $\frac52^+$ (top) and
excited $\frac12^+$ (bottom) states.
Predictions obtained using several versions of nonlocal optical potentials of type (\ref{eq:Vpb}) and (\ref{eq:Vdj}),
displayed by solid and dashed-dotted curves, respectively,
and those using the local optical potential from  Ref.~\cite{kobos:79} (dotted curves),
are compared with the experimental data from Ref.~\cite{dO25-63}.}
\end{figure}

 In summary, all three discussed nonlocal potential versions are phenomenological, become local in the limit
 $\rho \to 0$, and their parameters have to be determined by fitting the two-body data.
 Here we compare the potentials of type (\ref{eq:Vpb}) and (\ref{eq:Vdj}) in the
 $p+n+{}^{16}\mathrm{O}$ system that has been extensively studied in earlier works without CeX.
 First, for the two-body scattering $p+{}^{16}\mathrm{O}$ and  $n+{}^{16}\mathrm{O}$ the reproduction
 of the experimental data is of comparable quality as achieved in Ref.~\cite{deltuva:16d} with
  the nonlocality type  (\ref{eq:Vpb})
(excluding backward angles that need special terms). Second, we also calculated 
  ${}^{16}\mathrm{O}(d,p){}^{17}\mathrm{O}$ reactions using several versions of
  type (\ref{eq:Vpb}) and (\ref{eq:Vdj})  nonlocal potentials, labeled in Fig.~\ref{fig:O} by nonlocal(16) and
  nonlocal(23), respectively. We show examples
  at the deuteron beam energy $E_d = 36$ MeV for transfer reactions leading either to the ${}^{17}\mathrm{O}$
  ground state $\frac52^+$ or   excited state $\frac12^+$. In addition, we include the results
  from Ref.~\cite{deltuva:16d} obtained with a local potential \cite{kobos:79}. 
  It is obvious that both types of the nonlocal potential predict very similar nonlocality effects,
  and are clearly superior in accounting for the experimental data \cite{dO25-63} as compared to the local one.

  Thus, the proposed nonlocal potential  (\ref{eq:Vdj}) captures well the essential features of a more
  traditional  nonlocal potential  (\ref{eq:Vpb}). The next step is its extension 
 to include the excitation of the core.
 The inclusion of the CeX for local potentials with the rotational model  \cite{tamura:cex,nunes:96a},
 as appropriate for ${}^{10}\mathrm{Be}$, assumes 
a permanent quadrupole deformation of the core nucleus characterized
by the deformation parameter $\beta_2$, resulting in the Woods-Saxon radius $R_k = R_{k0}[1+\beta_2 Y_{20}(\hat{\xi})]$ 
depending on the internal core degrees of  freedom $\hat{\xi}$ in the body-fixed frame.
This renders the   potential noncentral; it is expanded into multipoles retaining  the $\lambda=2$ multipole
as a dominant one for the CeX.
The new nonlocal form (\ref{eq:Vdj}) has an advantage in performing partial-wave and multipole expansion
\begin{equation}  \label{eq:Vdjl}
  \begin{split}
    \langle r'L'S'J| V_{N}^{ba} |rLSJ \rangle = {} & \frac12 \Big[ H_{L'}(r',r) V_{L'S',LS,J}^{ba}(r) \\ & +
    V_{L'S',LS,J}^{ba}(r') H_{L}(r',r) \Big]
  \end{split}
\end{equation}
since the respective integration angular variables  are not coupled,
and consequently the terms $H(|\mbf{r}'-\mbf{r}|)$ and $V(r)$ can
be expanded separately, resulting in $H_{L}(r',r)$ and $V_{L'S',LS,J}^{ba}(r)$,
with orbital momentum $L$, total spin $S$ and conserved total angular momentum $J$.
$V_{L'S',LS,J}^{ba}(r)$ is exactly the  standard local potential with the CeX
\cite{tamura:cex,nunes:96a}, employed also in previous Faddeev-type calculations.
The transformation of the potential (\ref{eq:Vdjl}) to the momentum space is straightforward.
Note that  for the nonlocal potential of type (\ref{eq:Vpb}) the angular variables of 
partial-wave and multipole expansion are coupled, since deformed $V(y)$ depends both on $\xi$ and the
angle between $\mbf{r}'$ and $\mbf{r}$. This precludes taking over directly the result derived for the local
potential.

\begin{figure}[!]
\begin{center}
\includegraphics[scale=0.56]{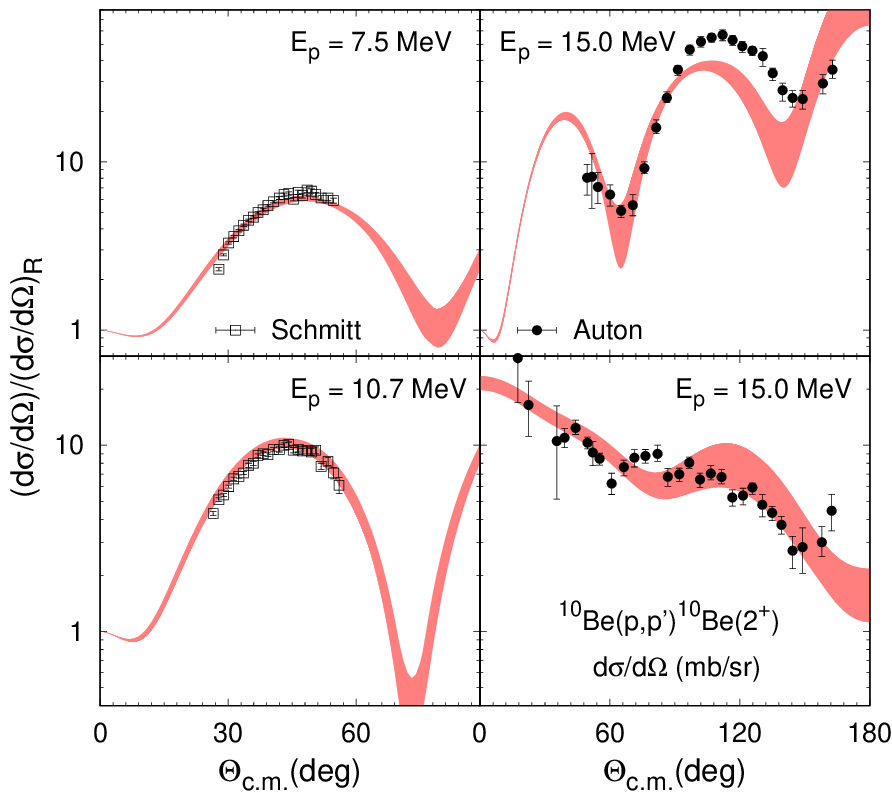}
\end{center}
\caption{\label{fig:p10be}  (Color online)
  Differential cross sections  $d\sigma/d\Omega$ divided by the Rutherford cross section for  elastic
  $p+\A{10}{Be}$ scattering at 7.5, 10.7 and 15 MeV/nucleon
  beam energies as functions of the c.m. scattering angle $\Theta_\cm$.
  The bottom-right panel shows the differential cross section for the
  inelastic   $p+\A{10}{Be}$ scattering at 15 MeV/nucleon energy.
  Predictions with different parameter sets of the nonlocal optical potential are combined into bands and
  compared with the experimental data from Refs.~\cite{dBe12-21c}
and \cite{dBe12p}.}
\end{figure}

The parameters of the new nonlocal potential for the  $p+{}^{10}\mathrm{Be}$ system are partly assumed and partly
determined by fitting the experimental data \cite{dBe12-21c}.
The Coulomb force and its deformation \cite{tamura:cex} is included as well.
We introduced some constraints before the fit such that our potential has no more free parameters than
standard optical potentials. First, we fix the nonlocality range $\rho=1$ fm, close to a typical value for nonlocal
potentials \cite{giannini,giannini2}. Second, as for several standard local potentials \cite{CH89,watson}
we assume the same geometric parameters for both imaginary terms, i.e., $f_W(y)=f_S(y)$. 
Third, since no polarization data are available to constrain the spin-orbit force,
we assume  $f_s(y)=f_V(y)$ and $V_s = 7.5\, \mathrm{MeV\,fm^2}$, close to the value for other
nonlocal potentials \cite{deltuva:16d}, and do not deform this term.
Thus, the parameters to be determined from the fit are the strengths
$V_V$, $W_V$, and $W_S$, reduced radii $r_V$ and $r_W$,  diffuseness $a_V$ and $a_W$, and the quadrupole
deformation parameter $\beta_2$. We fitted them to the elastic data at 6, 7.5, 9, 10.7  MeV per nucleon
\cite{dBe12-21c} and to both elastic and inelastic data at 12, 13, 14, 15 and 16 MeV per nucleon
\cite{dBe12p}.
In order to estimate the uncertainties we developed about 10 parameter sets with a comparable quality of the fit,
and show the corresponding predictions  as  bands in Fig.~\ref{fig:p10be}.
An example set is $V_V = 91.17$ MeV, $W_V = 1.35$ MeV,  $W_S = 7.61$ MeV,
$r_V = 1.14$ fm, $r_W = 1.26$ fm, $a_V = 0.50$ fm, $a_W = 0.65$ fm, and $\beta_2$ = 0.72;
the parameter tables can be obtained from the authors upon request. The value of the quadrupole
deformation parameter $\beta_2$ ranges between 0.71 and 0.78, quite consistent with 0.70 to 0.74 
from the DWBA analysis \cite{dBe12p} using local potentials. 
We stress that we achieve 
quite satisfactory description of the data  by using energy-independent potentials.

Regarding the $n+{}^{10}\mathrm{Be}$ potential we are not aware of the available experimental data. We
therefore fix all the geometry parameters to those of the $p+{}^{10}\mathrm{Be}$ potential,
as done for most of the local potentials, and refit only the strengths $V_V$, $W_V$, and $W_S$,
aiming to make the predictions consistent with those of standard local optical potentials.
The achieved consistency is comparable to the one between  the $p+{}^{10}\mathrm{Be}$ predictions
and data, and is not shown separately.

\section{Results \label{sec:res}}

In this section we present results for ${}^{10}\mathrm{Be}(d,p){}^{11}\mathrm{Be}$ and 
${}^{11}\mathrm{Be}(p,d){}^{10}\mathrm{Be}$ reactions obtained solving the Faddeev-type three-body equations
and including the quadrupole excitation of the ${}^{10}\mathrm{Be}$ core via the
potentials developed in the previous section. In the $n+{}^{10}\mathrm{Be}$ partial waves  with total
angular momentum/parity $\frac12^+$ ($\frac12^-$) where the   ground (excited) state of  $\A{11}{Be}$ resides we use
real nonlocal potential with geometric parameters taken over from Refs.~\cite{deltuva:13d,nunes:96a}
and fit the strengths to reproduce the binding energy of 0.504 MeV (0.184 MeV) for the ground (excited) state.
In the  $\frac12^+$ case we add a weak (few percent of the central volume part) $\mbf{L}^2$ term
as in Refs.~\cite{deltuva:17b,amos:03a}
adjusted to reproduce the desired binding energy and the SF. 
Finally, the realistic CD Bonn potential \cite{machleidt:01a}
is used for the neutron-proton interaction, and the $p+{}^{10}\mathrm{Be}$ Coulomb force is included
via the screening and renormalization method \cite{taylor:74a,taylor:74b,alt:80a,deltuva:05a}
as in Refs.~\cite{deltuva:13d,deltuva:17b}.

\begin{figure}[!]
\begin{center}
\includegraphics[scale=0.6]{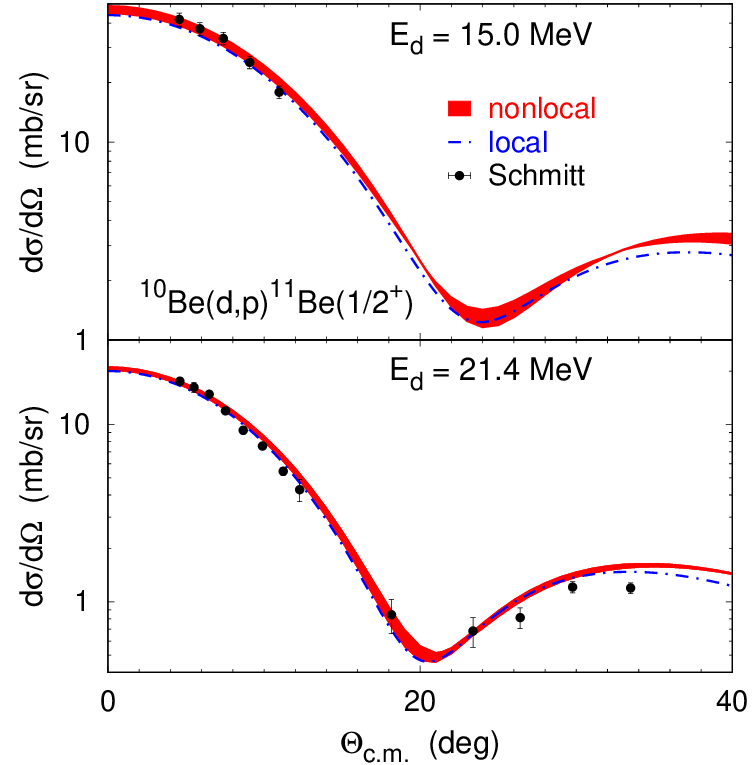}
\end{center}
\caption{\label{fig:s15-21}
Differential cross section for the transfer reaction 
$\A{10}{Be}(d,p)\A{11}{Be}$  at 15 and 21.4 MeV deuteron beam energy, the final $\A{11}{Be}$ nucleus
being in its ground state $\frac12^+$.
Predictions with the spectroscopic factor $S(0^+,\frac12^+)=0.754$ and 
different parameter sets of  nonlocal optical potentials are combined into a band and
  compared with the experimental data from Ref.~\cite{dBe12-21c}.
}
\end{figure}

We start with the study of the $\A{10}{Be}(d,p)\A{11}{Be}$ reaction at deuteron beam energies
$E_d = 15$ and 21.4 MeV
where the final  $\A{11}{Be}$ nucleus is in its ground state $\frac12^+$.
We performed calculations using different parameter sets of $p+{}^{10}\mathrm{Be}$ and  $n+{}^{10}\mathrm{Be}$
nonlocal optical potentials (\ref{eq:Vdjl}) and combined them into a band.
The transfer cross section depends also on the spectroscopic factor $S(S_A^{\pi_A},J^{\Pi})$ where
$S_A^{\pi_A}$ and $J^{\Pi}$ are spin/parity of  $\A{10}{Be}$ and $\A{11}{Be}$, respectively.
Since for small variations of the SF that dependence is nearly linear,
in Fig.~\ref{fig:s15-21} we show the predictions for  a single fixed  $S(0^+,\frac12^+)=0.754$ value that
is found to be consistent with the experimental data \cite{dBe12-21c}.
The width of the band reflects the theoretical uncertainty due to the potential model,
and therefore also the theoretical error bar for the SF, which is around 5\% (3\%) at 
15 MeV (21 MeV).
Remarkably, the predictions from Ref.~\cite{deltuva:15b} based on the local CH89 potential with
$S(0^+,\frac12^+)=0.855$ and displayed by dashed-dotted curves turn out to be very close.
For comparison, the SFs extracted in Ref.~\cite{dBe12-21c} using ADWA with CH89
and KD potentials \cite{CH89,koning}, range from 0.77 to 0.81 and from 0.67 to 0.74
at $E_d = 15$ and 21.4 MeV, respectively. Thus, our results show considerably weaker energy-dependence.

\begin{figure}[!]
\begin{center}
\includegraphics[scale=0.6]{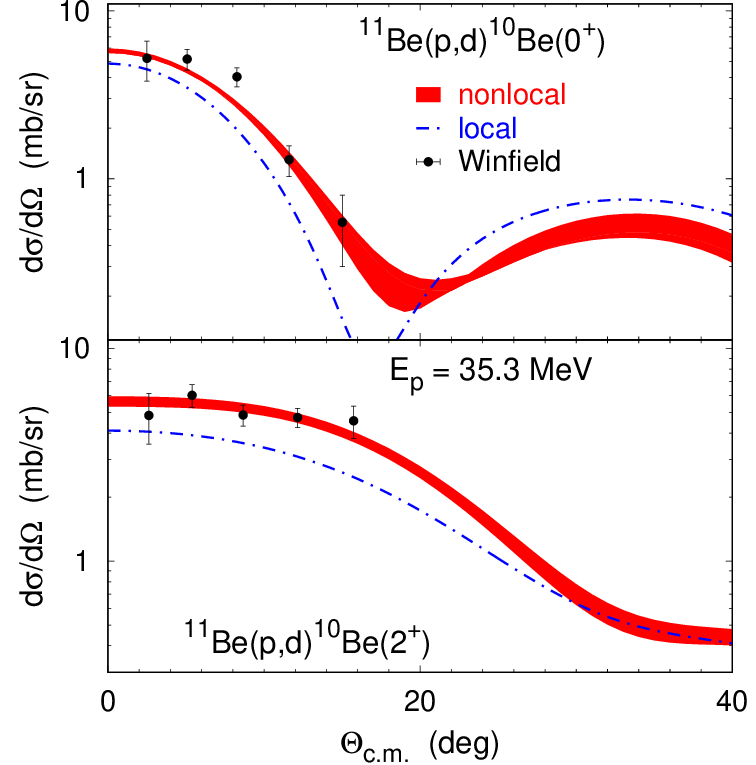}
\end{center}
\caption{\label{fig:s35} (Color online)
Differential cross section for
$\A{11}{Be}(p,d)\A{10}{Be}$  transfer reactions at $E_p = 35.3$ MeV
leading to the ground ($0^+$) and excited ($2^+$) states of $\A{10}{Be}$.
Predictions with  spectroscopic factors $S(0^+,\frac12^+)=0.807$, $S(2^+,\frac12^+)=0.193$ and 
different parameter sets of  nonlocal optical potentials are combined into a band and
  compared with the experimental data are from Ref.~\cite{winfield:01}. }
\end{figure}

The above agreement is likely accidental as it disappears at higher energy. 
In  Fig.~\ref{fig:s35} we study the deuteron pickup reaction $\A{11}{Be}(p,d)\A{10}{Be}$
at $E_p = 35.3$ MeV/nucleon beam energy with the  final $\A{10}{Be}$ nucleus being in its ground (excited)
state $0^+$ ($2^+$). This energy is equivalent to $E_d = 40.9$ MeV in the time-reversed
$\A{10}{Be}(d,p)\A{11}{Be}$  reaction.
Again,  predictions with different parameter sets of nucleon-nucleus
nonlocal optical potentials (\ref{eq:Vdjl}) are combined  into a band, but this time a better agreement
with the experimental data \cite{winfield:01} is obtained with ~5\% change in the SF,
i.e.,  $S(0^+,\frac12^+)=0.807$ and, consequently,  $S(2^+,\frac12^+)=0.193$,
which quite reasonably compare with  0.90 and 0.16, respectively, as predicted by the
ab initio no-core shell model with continuum (NCSMC) with an extra adjustment to neutron separation energies \cite{calci:16a}.
The uncertainties associated with the band width are around 3\% and 5\%, respectively.
With this model and nonlocal potentials we are able to provide rather good description of the experimental data
simultaneously for both transfer reactions leading to either ground or excited state of $\A{10}{Be}$.
In contrast,  the local potential predictions from  Ref.~\cite{deltuva:15b}
underestimate the differential cross section in both cases,
and the description can not be repaired by an adjustment of SFs (assuming their sum is one),
since increasing the cross section for one channel would reduce it for another.
As shown in Ref.~\cite{deltuva:15b} none of the  three different local optical potentials
was able to reproduce the data \cite{winfield:01} well. DWBA analysis \cite{winfield:01}
reports $S(0^+,\frac12^+)$ values from 0.65 to 0.80.

\begin{figure}[!]
\begin{center}
\includegraphics[scale=0.6]{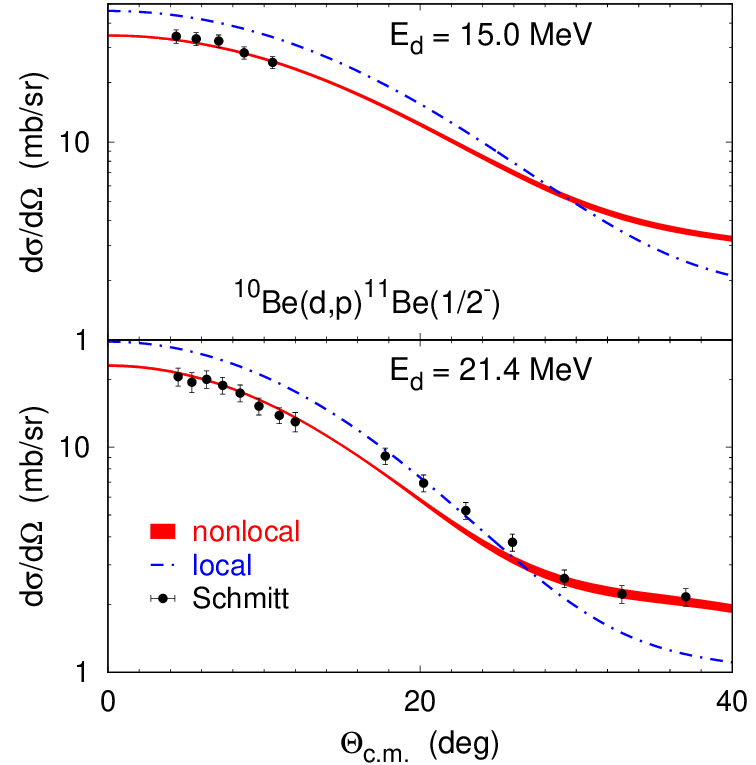}
\end{center}
\caption{\label{fig:p15-21}
Differential cross section for transfer reactions 
$\A{10}{Be}(d,p)\A{11}{Be}$  at $E_d = 15$ and 21.4 MeV
leading to the excited $\frac12^-$ state of $\A{11}{Be}$.
Predictions with the spectroscopic factor $S(0^+,\frac12^-)=0.654$ and 
different parameter sets of  nonlocal optical potentials are combined into a band and
  compared with the experimental data from Ref.~\cite{dBe12-21c}.
}
\end{figure}

Finally, Figure \ref{fig:p15-21} presents our results for the differential cross section
in the  $\A{10}{Be}(d,p)\A{11}{Be}$ reaction at $E_d = 15$ and 21.4 MeV but with the final  $\A{11}{Be}$ nucleus
being in its excited state $\frac12^-$. In this partial wave we used a real nonlocal potential that 
yields  $S(0^+,\frac12^-)=0.654$, while predictions with different parameter sets of nucleon-nucleus
nonlocal optical potentials (\ref{eq:Vdjl}) form  a band. At a first glance it seems to be  narrower than the
the $p+{}^{10}\mathrm{Be}$ band in Fig.~\ref{fig:p10be}, however, the width of bands for
angles up to 40 deg is quite comparable.
With $S(0^+,\frac12^-)=0.654$ the nonlocal potential predictions describe the experimental data
\cite{dBe12-21c} quite well at both energies, where the error bar associated with the band width is below 1\%.
 The local potential results taken
from  Ref.~\cite{deltuva:15b} correspond to  $S(0^+,\frac12^-)=0.786$ and fail to reproduce the data.
The agreement for forward angles could be improved by rescaling the predictions 
with smaller  $S(0^+,\frac12^-)\approx 0.60$ instead, but the disagreement at $\Theta_\cm > 20$ deg would increase.
There is an important difference in the shape of the angular distribution obtained with local and nonlocal potentials,
consistent with findings of  earlier works without the CeX. 
Our $S(0^+,\frac12^-)=0.654$ value is in agreement with ADWA results \cite{dBe12-21c}
ranging from 0.63 to 0.71.

We finally note that the present calculations cannot explain systematically lower cross sections
at $E_d=18$ MeV \cite{dBe12-21c}, a problem present also in all earlier ADWA and Faddeev-type calculations
 \cite{deltuva:13d,dBe12-21c}.

Since the nucleon transfer reactions at low energies and forward angles are peripheral to a good approximation,
the differential cross sections should scale with the square of the asymptotic normalization coefficient (ANC)
\cite{yang:18a}. We therefore collect in Table \ref{tab:anc} the ANC values 
corresponding to the ${}^{11}\mathrm{Be}$ wave functions used in this work, in an earlier Faddeev-type study
with local potentials \cite{deltuva:15b}, the ones extracted in the ADWA study \cite{yang:18a}
with the CH89 optical potential, and 
those resulting from the NCSMC calculations \cite{calci:16a}.
Where available, we include  in Table \ref{tab:anc} also SF values. Despite differences in dynamics and SF values,
 we observe a good agreement
for the ground state ANC. The excited state ANC obtained with the nonlocal model deviates from others by ~10\%,
but despite this fits the experimental data at forward angles as good as  other models, an provides even
a better description at larger angles. This indicates that beside the ANC also the continuum dynamics plays
important role in transfer reactions.

\begin{table} [!]
  \caption{\label{tab:anc}
    Spectroscopic factors and asymptotic normalization coefficients (in units of $\fm^{-1/2}$)
    for ground and excited states of ${}^{11}\mathrm{Be}$  obtained by several reaction and structure models.
    The ANC values corresponding to  Figs.~\ref{fig:s15-21} and \ref{fig:s35} are
    0.779 and 0.809 $\fm^{-1/2}$, respectively.
    The results of Ref.~\cite{deltuva:15b} for the excited state were rescaled to fit the data at forward angles.
  }
\setlength{\tabcolsep}{5pt}
\begin{tabularx}{\columnwidth}{l|*{2}{c}}
\hline
& SF &  ANC    \\ \hline
 $S_A^{\pi_A} = 0^+, \quad J^{\Pi}=\frac12^+$ \\
 Faddeev, nonlocal & $0.78\pm0.04$ & $0.79\pm0.02$ \\
Faddeev, local \cite{deltuva:15b} & 0.854 & 0.785 \\
ADWA \cite{yang:18a} &  & $0.785 \pm 0.03 $  \\ 
NCSMC  \cite{calci:16a} & 0.90 & 0.786 \\ \hline
$S_A^{\pi_A} = 0^+, \quad J^{\Pi}=\frac12^-$ \\
Faddeev, nonlocal & $0.65\pm0.01$ & $0.149\pm0.005$ \\
Faddeev, local \cite{deltuva:15b} & 0.60 & 0.13 \\
ADWA \cite{yang:18a} &  & $0.135 \pm 0.005 $  \\ 
NCSMC  \cite{calci:16a} &  & 0.129 \\ 
\hline
\end{tabularx}
\end{table}

\section{Conclusions \label{sec:sum}}

We proposed a new nonlocal form of the nucleon-nucleus optical potential,
and using the ${}^{16}\mathrm{O}(d,p){}^{17}\mathrm{O}$ reaction as example
demonstrated that it reproduces well the essential features of traditional nonlocal potential.
We extended the new nonlocal potential to account for collective degrees of freedom of the nuclear core
via the rotational quadrupole excitation.
For the $p+{}^{10}\mathrm{Be}$ system the potential parameters were determined
by fitting the experimental data for elastic and inelastic scattering.
In contrast to standard local potentials,
developed  nonlocal parametrizations are energy-independent but nevertheless are able to  provide
quite a good description of the experimental data over a broader energy range.
For the $n+{}^{10}\mathrm{Be}$ system, owing to the lack of the experimental data,  the potential parameters
were determined demanding consistency with the predictions by several global parametrizations of standard optical potentials.

Nonlocal potentials with explicit excitation of the core for the first time were applied
to the study of deuteron stripping and pickup reactions using rigorous three-body scattering theory.
Faddeev-type equations extended for the core excitation  were solved in the momentum-space
partial-wave representation leading to well-converged results.
We studied deuteron stripping and pickup reactions
${}^{10}\mathrm{Be}(d,p){}^{11}\mathrm{Be}$ and ${}^{11}\mathrm{Be}(p,d){}^{10}\mathrm{Be}$
at energies corresponding to 15, 21.4 and 40.9 MeV deuteron beam energy.
A good description of the experimental data \cite{dBe12-21c,winfield:01} was achieved,
though for the reactions involving the ${}^{11}\mathrm{Be}$ ground state a slight
readjustment of the SF  by $\sim 5$\% was required at the highest energy.
On the other hand, the predictions using local optical potentials have been less successful.
The SF values found to be consistent with the  experimental data are
$S(0^+,\frac12^+) = 0.78 \pm 0.04$  and $S(0^+,\frac12^-) = 0.65 \pm 0.01$,
in reasonable agreement with some of earlier determinations \cite{dBe12-21c,winfield:01} based on DWBA or ADWA.
However, an important feature of our results is smaller spread of values and weaker energy dependence,
likely due to more sophisticated potentials and treatment of the three-body dynamics.
The results also suggest that with the nonlocal potential parameters adjusted to the
experimental data in a limited energy regime it may be possible to make reliable predictions outside
that regime, indicating an increased predicting power of our calculational scheme.
This is expected to hold also for other reaction channels such as inelastic and breakup, and the 
future studies should clarify this question.

\vspace{1mm}

This work was supported by Lietuvos Mokslo Taryba
(Research Council of Lithuania) under Contract No.~S-MIP-22-72.



\end{document}